\documentclass[usegraphicx,useAMS,usenatbib]{mn2e}
\usepackage{longtable} 
\usepackage{longtable} 
\usepackage{rotating}

\usepackage{times}

\title[Starspots on M-dwarfs in NGC~2516]
 {Why do some young cool stars show spot modulation while others do not?
}

\author[R. J. Jackson and R. D. Jeffries]
  {R. J.~Jackson and R. D.~Jeffries\\
   Astrophysics Group, Research Institute for the Environment, Physical Sciences and Applied Mathematics, Keele University,\\
   $\;$ Keele, Staffordshire ST5 5BG, UK}

\date{Accepted for publication in MNRAS}

\pagerange{\pageref{firstpage}--\pageref{lastpage}} \pubyear{2011}

\def\LaTeX{L\kern-.36em\raise.3ex\hbox{a}\kern-.15em
    T\kern-.1667em\lower.7ex\hbox{E}\kern-.125emX}

\begin{document}
\label{firstpage}
\maketitle

\begin{abstract}
We present far-red, intermediate resolution spectroscopy of 572
photometrically selected, low-mass stars ($0.2<M/M_{\odot}<0.7$) in the
young open cluster NGC~2516, using the FLAMES spectrograph at the Very
Large Telescope. Precise radial velocities confirm membership
for 210 stars that have published rotation periods from
spot-modulated light curves and for another 144
stars in which periodic modulation could not be found. The two
sub-samples are compared and no significant differences are found
between their positions in colour-magnitude diagrams, the distribution
of their projected equatorial velocities or their levels of
chromospheric activity.
We rule out differing observational sensitivity
as an explanation and conclude that otherwise similar objects, with equally
high levels of chromospheric activity, do not exhibit spot-induced
light curve modulation because their significant spot coverage is
highly axisymmetric. We propose that the spot coverage consists of
large numbers of small, dark spots with diameters of about
$2^{\circ}$. This explains why about half of cluster members do not
exhibit rotationally modulated light curves and why the light curve
amplitudes of those that do have mean values of only 0.01-0.02 mag.

\end{abstract}

\section{Introduction}

Drawing on a solar analogy, intense dynamo-generated magnetic fields
are thought to be responsible for regions of suppressed convective flux
and cooler temperatures in late-type stellar photospheres
(see Thomas \& Weiss 2008 and Strassmeier 2009).  
Evidence for dark ``starspots'' and their associated magnetic
fields initially came from rotationally modulated 
broadband fluxes (Hall 1972, Eaton \& Hall 1979 and Vogt 1981), but
there is now a large literature that detects and investigates starspots
using direct and indirect observational techniques -- Doppler
imaging (Collier-Cameron \& Unruh 1994, Rice, Strassmeier \& Linsky
1996, Strassmeier 2002), measurements of the flux in molecular bands
(e.g. Neff, O'Neal \& Saar 1995, O'Neal et al. 2004), spectroscopy of
Zeeman-broadened lines (e.g. Marcy 1982, Gray 1984, Johns-Krull \&
Valenti 1996) and tomography using circularly polarized light (Zeeman
Doppler Imaging -- Semel 1989, Donati et al. 1997).

The size, filling factor and position of spots as a function of time
and as a function of stellar mass and rotation rate are important
probes of the poorly understood dynamo mechanism, sometimes in physical
circumstances quite unlike the Sun; for example, in stars that are very
fast rotators or are fully convective. Understanding the properties of
starspots is also important because they and their associated magnetic
fields may be responsible for increasing the radii of low-mass stars
and lowering their effective temperatures (Chabrier, Gallardo \&
Baraffe 2007, Ribas et al. 2008, Jackson, Jeffries \& Maxted
2009). Starspots complicate the interpretation of eclipsing binary and
transiting planet light curves and compromise the accuracy with which
fundamental stellar and planetary radii can be determined (Watson \&
Dhillon 2004, Czesla et al. 2009, Morales et al. 2010). 
Photospheric starspots provide a nuisance source of systematic
uncertainty in precision astrometric and radial velocity measurements,
which may limit the detection of low-mass planets (Makarov et al. 2009,
Barnes, Jeffers \& Jones 2011).

Starspots are often observed on cool members of young open
clusters, usually as a means of estimating rotation periods via the
rotational modulation of light curves as spots transit the
visible hemisphere (van Leeuwen et al. 1987, Prosser et al. 1993).  A
review of the literature is provided by Irwin \& Bouvier (2009), with
more recent examples including work by Irwin et al. (2009), Hartman et
al. (2010), James et al. (2010), Delorme et al. (2011) and Meibom et
al. (2011).  Many young stars in these clusters are rapid rotators with
vigorous dynamo action, leading to extensive optical, ultraviolet and
X-ray manifestations of the consequent magnetic field. Spot-induced
modulation is seen with a wide range of amplitudes from mmag levels up
to modulations larger than 0.2 mag (peak-to-peak). Some of this
diversity appears related to the expected connection between
dynamo-related phenomena and rotation rate. A strong correlation has
been found between modulation amplitudes and increasing
rotation rate or decreasing Rossby number (the dimensionless ratio of
rotation period to convective turnover time; O'Dell et
al. 1995, Messina 2001).  Nevertheless, even at a given rotation
rate or Rossby number there is still a wide range of light curve
variability. Sometimes rotational modulation is seen in a
large fraction of studied cluster members. For example, Irwin et
al. (2008) found rotation periods for about 50 per cent of members in
the young (35\,Myr) cluster NGC 2547, and Hartman et al. (2010) found
rotation periods for 74 per cent of low-mass Pleiades (age 120\,Myr)
members. However, in other studies this fraction can be very much
lower. A study of low-mass stars in Praesepe (age
600\,Myr) found rotation periods for only
$\sim 10$ per cent of members (Ag\"{u}eros et al. 2011).

Stars where rotation periods cannot be measured may offer important
clues to the dynamo mechanism or the surface distribution of starspots,
and their dependence on mass and rotation rate.  Alternatively, the
non-detections may simply be a result of observational
biases in sensitivity and sampling.  When using measured rotation
periods to investigate angular momentum evolution it is important to
understand whether the observed samples with detected rotation rates
are biased according to their rotation rates. For example, slower
rotators may have smaller spot-induced light curve variations that
evade detection.

In this paper we take advantage of a homogeneous survey for rotation
periods undertaken in the young (150\,Myr) open cluster NGC~2516 by
Irwin et al. (2007, hereafter IHA07).  In previous work (Jackson et
al. 2009, Jackson and Jeffries 2010) we presented spectroscopic
observations of the low-mass ($0.2<M/M_{\odot}<0.7$) stars in NGC~2516
{\it with} detected rotational modulation and period estimates from
IHA07. We measured projected equatorial velocities and chromospheric
activity and showed that these rapidly rotating stars were highly
magnetically active and likely to have large filling factors of dark
starspots. Here we extend this study to a large sample of photometric
candidates in NGC~2516 that {\it did not} have measured periods in an
effort to understand why some low-mass stars exhibit rotational light
curve modulation and some do not.

Section 2 describes the spectroscopy and determination of
radial velocity (RV), projected equatorial velocity ($v\sin i$) and
chromospheric activity. In section 3 the RV data are used to
determine a list of cluster members and measure the proportion of stars
with and without measured rotational periods. In section 4 we compare
the measured properties of these two groups of stars.  Finally, in
section 5 we discuss what form of spot distribution could give rise to
the observed properties of fast rotating, low-mass stars in
NGC~2516 and how this distribution might affect other methods of
observing starspots on young stars.

\section{Spectroscopic observations of M-dwarfs in open cluster NGC~2516}

NGC~2516 is a relatively close and well studied, young open
cluster. Recent papers describe membership surveys and
characterisation of the cluster mass function (Jeffries, Thurston \&
Hambly 2001; Sung et al. 2002; Moraux, Bouvier \& Clark 2005). The age
of the cluster has been determined as $\simeq$150~Myr from the nuclear
turn off in high mass stars and the lithium depletion and X-ray
activity seen in cooler stars (Jeffries, James \& Thurston 1998; Lyra
et al. 2006). Metallicity is close to solar; determined
spectroscopically as [Fe/H]\,$= 0.01 \pm 0.07$ and photometrically as
[M/H]=\,$-0.05 \pm 0.14$ (Terndrup et al. 2002). The same authors give
an intrinsic distance modulus of $7.93 \pm0.14$ ($385\pm25$~pc) based
on main sequence fitting and a cluster reddening of $E(B-V) = 0.12 \pm
0.02$. These values of distance modulus and reddening are used in what
follows. Previous observations of chromospheric and coronal activity of
late K and M-dwarfs in NGC~2516 are summarised in Jackson and Jeffries
(2010, hereafter JJ10).

IHA07, surveyed NGC~2516 as part of a larger monitoring survey of young
open clusters (Hodgkin et al. 2006; Aigrain et al. 2007). They
identified about 1000 potential cluster members from a $V$ versus $V-I$
colour magnitude diagram (CMD), selecting targets with apparent
magnitudes in the range $16<V<26$ that are close to an empirically
defined isochrone. Periods were reported for 362 candidate members in the mass
range 0.15--0.7 $M_\odot$.

\subsection{Target selection and observations}

Targets for our spectroscopic survey were selected from IHA07 with
$14<I<19$.  They were observed using the European Southern Observatory
(ESO) 8.2m aperture Very Large Telescope (UT-2 Kueyen) FLAMES fibre
instrument, feeding the Giraffe and UVES spectrographs. Multiple
targets were observed with Giraffe in eight separate fibre
configurations. The allocation of fibres to targets was biased in the
sense that targets with measured periods in IHA07 were placed first and
the remaining fibers allocated to candidate members from IHA07
without measured periods. About 15 fibres in each configuration were
reserved for ``blank'' sky positions.

The Giraffe spectrograph was used with the HR20A grating, covering 
wavelengths 8060-8600\AA\ at a resolving power of 16000. Bright,
early type stars in NGC~2516 were selected from the catalogue of Dachs
and Kabus (1989) for use as telluric standards. Several of these stars
in each configuration were observed with UVES at a resolving power of
47000. Details of the eight configurations are shown in Table~1 and
their spatial locations indicated in Fig.~1. Each configuration was
observed with two sequential 1280\,s Giraffe exposures and three
sequential 800\,s UVES exposures.  One configuration was observed twice
during the same night. The repeated data were useful in assessing
measurement uncertainties.  A total of 981 spectra were recorded for
824 unique targets.

\begin{table}
	\caption[{Details of observing programme 380.D-0479}]{Details
	of observing programme 380.D-0479}
\begin{tabular}{cllccc}
		\hline
		Ref No. & Date \& & Centre & Seeing & No. of & with\\
		Run No. & start time & RA / Dec & start/end & targets  & period  \\\hline
		 287508 & 27-11-07 & 118.639  & 0.78& 108 (108) & 45 (45)\\
		 1& 04:57 & -60.804 & 0.79 & \\
		 287510 & 27-11-07 & 118.209 & 0.91 &  103 (101) & 22 (21)\\
		 2& 05:53 & -61.107 &  0.75 & \\
		 287512 & 27-11-07 & 119.102  & 0.69 &  112 (110) & 44 (43)\\
		 3& 06:48 & -61.108 &  0.79 & \\
		 repeat & 27-11-07 & 119.102  & 0.91 &  112 (0) & 44 (0)\\
		 4& 07:39 & -61.108  & 0.89 & \\
		 287514 & 02-01-08 & 120.010  & 1.15 &  110 (110) & 38 (38)\\
		 5& 06:25 & -61.265  & 1.03 & \\
		 287516 & 29-11-07 & 119.804 &  1.19 &  111 (104) & 55 (50)\\
		 6& 06:51 & -60.934 & 1.30 &  \\
		 287518 & 02-01-08 & 120.594  & 1.15 &  102 (102) & 41 (41)\\
		 7& 05.31 & -60.790  & 1.16 & \\
		 287520 & 30-12-07 & 120.097  & 0.94 &  112 (110) & 39 (38)\\
		 8& 02:40 & -60.549  & 0.85 & \\
		 287522 & 30-12-07 & 119.462  & 1.21 &  111 (83) & 32 (22)\\
		 9& 01:44 & -61.402  & 1.21 & \\ \hline
		 \multicolumn {6} {l} {The field centres are in degrees and the seeing in arcsec.}\\
		 \multicolumn {6} {l} {Values in brackets indicate the number of new targets observed per OB.}\\
		 \end{tabular}
	\label{table1}
\end{table}

\begin{figure}
	\centering
		\includegraphics[width = 84mm]{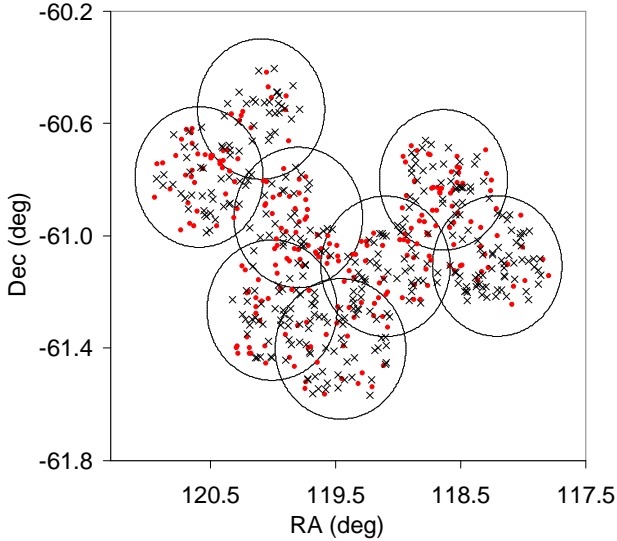}
\caption[Coordinates of targets in the open cluster
NGC~2516]{The distribution of targets in NGC~2516 (see
Table~1). Circles define the boundaries of the eight
FLAMES fields.  Filled circles show target stars with a
spectral signal-to-noise ratio (SNR) $\geq$ 5, and measured period.
Crosses show targets with a SNR $\geq$ 5 but no
period.}
\label{fig1}
\end{figure}

\subsection{Radial and projected equatorial velocities}
Spectra were optimally extracted and analysed using the pipeline
described in detail by JJ10.  RVs and $v\sin i$ were determined by
convolving the spectra of targets with those of standard stars over the
wavelength range 8061\AA~to 8530\AA, excluding the chromospherically
contaminated Ca\,{\sc ii} infra-red triplet (Ca\,{\sc ii}
IRT) components at 8498\AA\ and 8542\AA. Templates of type K4.5V
(HD~209100) and M6V (HD~34055) from the UVES atlas (Bagnulo et
al. 2003) were used, bracketing the range of target spectral types. For
spectra with a signal-to-noise ratio per extracted pixel (SNR) $\geq$5,
there was usually a clearly defined single peak in the cross
correlation function (CCF) giving an unambiguous value of RV and $v
\sin i$. For spectra with SNR$<$5, the CCF peak was often distorted by
random noise and these results were considered unreliable and
rejected from the sample. Spectra with SNR $\geq$5 were obtained for
588 stars. Sixteen of these still showed distorted CCF peaks that did
not yield an RV or $v\sin i$ (see Table 2).

\begin{table}
	\caption[Targets with SNR $geq$5 and unresolved
	RV]{Categorisation of targets with SNR $\geq$5 and unresolved
	RV. The co-ordinates and $I$ magnitudes are from IHA07, with $I$ corrected as described in section~2.4.}
		\begin{tabular}{llccl}
\hline
RA & Dec & Apparent & SNR		& Category \\
(J2000) & (J2000) & $I$ mag.&  & (see notes)\\
\hline
7 56 54.04 &	-61 29 59.48 &	15.08	 & 28	&	Broad peak\\
7 58 39.38 &  -61 33 40.48 &	17.48	 & 5	&	Possible binary\\
7 56 22.44 &	-61 21 29.30 &	17.87	 & 5	&Low SNR\\
7 56 28.62 &	-61 17 11.80 &  15.00  & 37	&Broad peak (p)\\
7 57 17.64 &  -61 15 31.00 &  17.85	 & 6	&	Low SNR\\
7 58 46.35 &  -61 14 31.95 &	17.01	 & 6	&	Low SNR\\
7 56 19.26 &	-61 07 37.25 &	18.35	 & 5	&	Low SNR\\
7 56 56.36 &	-61 00 57.55 &  14.77	 & 44	&	Broad peak\\
8 00 58.89 &	-61 25 04.30 &  16.95  &12	&	Probable binary\\
7 59 55.73 &	-61 08 08.80 &	15.76	 & 29	&	Broad peak\\
7 59 32.60 &  -60 59 07.50 &  15.36	 & 26	&	Broad peak (p)\\
8 03 11.28 &	-60 34 22.24 &	15.57	 & 22	&	Broad peak\\
7 52 05.85 &	-60 56 55.33 &	18.28	 & 5	&	Possible binary\\
7 53 32.20 &	-61 07 55.22 &  15.17	 & 36	&	Broad peak\\
7 54 6.64	 &  -60 41 09.90 &  15.71	 & 27	&	Probable binary (p)\\
7 54 23.11 &  -60 44 34.46 &	15.65	 & 18	&	Probable binary\\\hline

\multicolumn {5} {l} { Probable binary- clear double peak in the CCF for high SNR.}\\
\multicolumn {5} {l} { Possible binary - apparent double peak in the CCF for low SNR.}\\		 
\multicolumn {5} {l} { Low SNR -  distorted peak in the CCF for low SNR.}\\		 
\multicolumn {5} {l} { Broad peak - CCF indicative of a fast rotator.  }\\		 
\multicolumn {5} {l} { (p) period reported in IHA07}\\

\end{tabular}
	\label{tab:Binaries}
\end{table} 

Uncertainties in RV and $v \sin i$ were determined
empirically by comparing repeated measurements for a subset of
targets (see JJ10 for details). This gave relationships
for uncertainties in RV
and CCF width, $W$, as a function of SNR and $v\sin i$;
 
\begin{equation}
\sigma _{\rm {RV}} = \sqrt{(9.2+0.013(v \sin i)^2)^2/{\rm SNR}^2 + 0.31^2} \ \ {\rm km\,s}^{-1}
\label{uncertainty_RV}
\end{equation}

\begin{equation}
\sigma _{W}  = \sqrt{(8.8+0.005(v \sin i)^2)^2/{\rm SNR}^2 + 0.40^2} \ \ {\rm km\,s}^{-1}
\label{uncertainty width}
\end{equation}

The uncertainty in $v\sin i$ was determined from $W$ and the ``zero
velocity width'', $K$ for each standard (where $K =19.1\pm 0.6$\,km\,s$^{-1}$
for the K4.5 standard and $K =17.7\pm 0.5$\,km\,s$^{-1}$ for the M6 standard).
Taking $v \sin i \propto (W-K)^{1/2}$ we obtain
\begin{equation}
\sigma_{v\sin i} = v\sin i \sqrt{\sigma_W^2 + \sigma_K^2} /[2(W-K)]\ \ {\rm km\,s}^{-1}.
\end{equation}

\subsection{Chromospheric activity indices}
The spectra include two of the Ca\,{\sc ii}~IRT lines with rest
wavelengths of 8498~\AA~and 8542~\AA.  These lines are known to be
effective chromospheric activity indicators, with chromospheric
emission filling the underlying absorption lines (Mallik 1994,
1997). The spectral subtraction technique used to measure the
equivalent width of the Ca\,{\sc ii}~IRT lines, follows that of
Marsden, Carter and Donati (2009) and is detailed in JJ10. The
conversion between equivalent width (EW) and activity index,
$R^{'}_{Ca}$, (the ratio of chromospheric Ca\,{\sc ii}~IRT flux,
corrected for any photospheric contribution, to the bolometric flux),
was established using continuum flux densities and colours tabulated by
Pickles (1998).

\begin{table*}
\caption{Velocity data measured for targets in NGC~2516.  The
identifier, co-ordinates and rotation period are taken from Table 1
of IHA07. $V$ and $I$ photometry are also from IHA07, but corrected in
the way described in section~2.4.  $K$-band photometry comes from
2MASS, transformed to the CIT system. Relative RV and $v \sin i$ values
are given for 572 targets which have a spectral SNR$\geq 5$, of which
354 are identified as cluster members. The right hand column indicates
the run number(s) for the observation (see Table 1), an asterisk
indicates a non-member.  The full table is available on Blackwell
Synergy as Supplementary Material to the on-line version of this
paper.}
\begin{tabular}{lllllllcccl} 
\hline 
Identifier & RA      & Dec     & Period & $V$     & $I_{C}$   & $K_{CIT}$ & SNR & RV     & $v \sin i$ & Run \\
    ~      & (J2000) & (J2000) & (d)    & (mag) & (mag) & (mag)   &  ~
    & (km\,s$^{-1}$) & (km\,s$^{-1}$)    &Nos. \\
 \hline       
N2516-1-1-1470 & 7 57 08.92 & -61 29 18.6 & ~8.803 & 17.60 & 15.45 & 13.47 & 26 & ~-0.56$\pm $0.47 &  ~$<$8.00  &  9 \\
N2516-1-1-1667 & 7 57 16.58 & -61 31 38.7 & ~1.347 & 15.82 & 14.50 & 12.72 & 44 & ~~0.16$\pm $0.46 &  21.98$\pm $1.91 &  9 \\
N2516-1-1-1880 & 7 57 24.78 & -61 25 56.6 & ~--- & 18.92 & 16.46 & 14.24 & 14 & ~-2.82$\pm $0.73 &  ~$<$8.00  &  9  * \\

 \end{tabular}
  \label{velocity_data}
\end{table*}

\begin{table*}
\caption{Chromospheric activity indices and related data for
targets in NGC~2516. The identifier, period and light curve amplitudes
are from IHA07. Luminosity, convective turnover time and Rossby number
are calculated as described in Section 2.4 (using $(M-m)_0=7.93$,
$A_V=0.37$, and $A_I=0.20$). EWs and chromospheric activity indices of
the first two Ca\,{\sc ii}~IRT lines (8498\AA\ and 8542\AA) are given
for 572 targets with resolved RV and SNR$\geq 5$. Quoted
uncertainties in EW and activity indices do not include possible
systematic errors (see JJ10). Masses are estimated
from the models of Baraffe et al. (1998) and the
absolute $I$ magnitude. The full table is available on Blackwell Synergy
as Supplementary Material to the on-line version of this paper.}
\begin{tabular}[t]{llccccccccl} \hline 
Identifier   & $\log_{10}$ & Period & sine & Turnover & $\log_{10}$& EW$_{8498}$& EW$_{8542}$ & log$_{10}$ & log$_{10}$&Mass \\
  & $L/L_{\odot}$ &  (days) & amplitude & time  & Rossby  & (\AA)  &   (\AA) & R'$_{Ca8498}$ & R'$_{Ca8542}$& $M_{\odot}$\\ 
           &    &    &(mag.)   & (days)  & No.   &          &           &             &            \\ \hline

N2516-1-1-1470 & -1.23 & 8.803 &.016 & ~52 & -0.77 & 0.45$\pm$0.02 & 0.50$\pm$0.03 & -4.47$\pm$.02 & -4.42$\pm$.02 & 0.57\\
N2516-1-1-1667 & -0.88 & 1.347 &.010 & ~35 & -1.41 & 0.42$\pm$0.01 & 0.59$\pm$0.01 & -4.49$\pm$.01 & -4.34$\pm$.01 & 0.70\\
N2516-1-1-1880 & -1.62 & --- &--- & ~81 & --- & 0.07$\pm$0.04 & 0.01$\pm$0.04 & -5.27$\pm$.03 & -6.18$\pm$.04 & 0.43\\
   \end{tabular}
  \label{chromospheric_data}
\end{table*}

\subsection{Tabulated results}
Results for the 572 targets with SNR $\geq$ 5 and a resolved RV are
given in Tables 3 and 4. In Table 3 the identifier, RA and Dec and period
(when measured) are taken from IHA07. The $V$ and $I$ magnitudes were
based on those given by IHA07, but corrected to put them onto the
better calibrated photometric scale of Jeffries et al.~(2001) using
stars common to both papers.  The corrections (added to the IHA07
values) were $\Delta I = 0.080-0.0076\,I$ and $\Delta (V-I) =
0.300-0.153\,(V-I)$.  $K$ magnitudes from the Two-micron All-Sky
Survey (2MASS) survey (Cutri et al. 2003) were transformed to the CIT
system using $K_{\rm CIT} = K_{\rm 2MASS} + 0.024$ (Carpenter 2001). No match
to the 2MASS catalogue was found for 4 targets. RVs are
shown relative to the average cluster RV, estimated using the RVs of
targets with $v \sin i < $20~km\,s$^{-1}$ (see section 3.1). 
A minimum resolvable value of 8~km\,s$^{-1}$
is taken for $v\sin i$. At this level of broadening the increase in
CCF width is $\simeq 1.3$ times the standard deviation of the
``zero~rotation'' width; hence there is a $\simeq 90$ percent
probability that $v\sin i$ is truly non-zero. For targets with repeated
measurements Table~3 shows weighted means of RV and $v\sin i$
with appropriate errors. 

Table 4 gives Ca\,{\sc ii}~IRT EWs and activity
indices. Also shown are the first harmonic amplitudes for stars with
rotational periods (from IHA07). The luminosity of the targets ($\log_{10}
L/L_{\odot} = 4.75-M_V - BC_V$) is calculated using a distance modulus
of $7.93$ (Terndrup et al. 2002) and bolometric corrections, $BC_V$,
interpolated as a function of $(V-I)_0$ using the tables of
Kenyon and Hartmann (1995). Colours were corrected assuming
a uniform reddening of $E(B-V)=0.12$ (Terndrup et al. 2002), ratios of
selective to total extinction $A_V/E(B-V) = 3.1$, $A_K/E(V-K)=0.13$
(Rieke \& Lebofsky 1985) and $E(V-I)=0.16$ (Jeffries et al. 2001). A 
stellar mass is estimated by interpolating the models of
Baraffe et al. (1998) using the absolute $I$ magnitudes.

Rossby numbers are given for stars with measured rotation periods, $P$,
calculated as $N_R = P/\tau_c$, where $\tau_c$ is the convective turnover
time. Rossby numbers are widely used to characterise the combined
effects of rotation and spectral type on the activity levels of F--K
stars. Their use for M-dwarfs is less clear since the widely-used
semi-empirical formula of Noyes et al.(1984) giving $\log \tau_c$ as a
function of $B-V$ is unconstrained for M-dwarfs.  An alternative
approach is followed here where the value of $\tau_c$ is chosen such
that chromospheric and coronal activity indicators satisfy a single
scaling law with Rossby number irrespective of stellar mass. Pizzolato
et al. (2003) noted that the mass dependence of a turnover time defined
in this way is closely reproduced by assuming $\tau_c \propto L_{\rm
bol}^{-1/2}$.  Anchoring the turnover time of a solar type star as
$\log \tau_c = 1.1$, this gives the following expression;

\begin{equation}
\log N_R = \log P - 1.1 + 0.5 \log(L_{\rm bol}/L_{\odot}).
\end{equation}

A few of the coolest stars in our sample (spectral type M4 and cooler,
$I\ga 18$) may be fully convective. The meaning of a turnover time in
these objects is ill-defined. We apply equation~4 to these
stars too, since others have shown that this
parameterisation describes the behaviour of rotation and activity in
stars at or just beyond the fully-convective boundary quite well. They
follow the same relationship between activity and Rossby number as
higher mass stars and their magnetic activity saturates at a similar
value of $N_R \simeq 0.1$ (Kiraga \& Stepien 2007; Jeffries et
al. 2011).

\section{Results}
\subsection{Membership}
Targets were identified as potential cluster members by IHA07 from
their position in the $V$/$V-I$ CMD. We make a
further selection based on RV relative to the mean RV of cluster
members in order to establish a more secure membership list.
Chromospheric activity and rotation were not used as indicators of
membership since we wish to study the distributions of these in cluster
members. Fig. 2 shows an RV histogram; the RVs of stars with measured
periods (shown as filled bars) are tightly bunched within a few km\,s$^{-1}$,
indicating that the majority of these are cluster members. Stars
without measured periods (shown as open bars) show a clear peak too but
a much broader spread, presumably due to the presence of a
proportion of contaminating older field stars.

\begin{figure}
	\centering
		\includegraphics[width = 84mm]{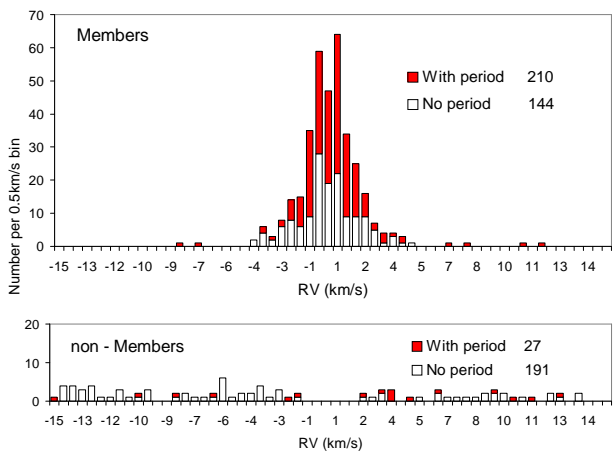}
	\caption[Number density of targets in NGC~2516 as a function of radial velocity]
	{Radial velocity histograms for targets in NGC~2516 (relative
	to the mean RV of the cluster). The upper plot shows objects we
	classify as cluster members in 0.5km\,s$^{-1}$ bins. The lower
	plot shows targets classified as non-members. Note that
	$\sim$50 percent of
	non members have RVs outside the range of this plot.}
	\label{fig2}
\end{figure}

\begin{table*}
\caption{A summary of target numbers and the fraction of cluster
  members with rotation periods as a function of $I$ magnitude}
\begin{tabular}{lcccccc}
\hline
 & Set 1 & Set 2 & Set 3 & Set 4 & Set 5 & All sets\\\hline
$I$ magnitude & 14.8 to 15.5 & 15.5 to 16.2 & 16.2 to 16.8 & 16.8 to 17.5
 & 17.5 to 18.2 & 14.8 to 18.2 \\
$M_I$ & 6.7 to 7.3 & 7.3 to 8.0 & 8.0 to 8.7 & 8.7 to 9.4
 & 9.4 to 10.1 & 6.7 to 10.1 \\
All targets (a)  & 80 +(4) & 99 +(2) & 129 +(0) & 116+(1) & 87+(2) & 511+(9)\\
Targets with period  & 38+(1) & 59+(0) & 59+(0) & 46+(0) & 25+(0) & 227+(1)\\
Percentage of targets with period (b) & 48$\%$ & 60$\%$ & 46$\%$ & 40$\%$ & 29$\%$ & 44$\%$\\
Percentage Monitor survey with period & 36$\%$ & 42$\%$ & 36$\%$ & 32$\%$ & 24$\%$ & 33$\%$\\
Targets which are members&50+(2) & 60+(1) & 81+(0) & 73+(1) & 67+(1) & 331+(5)\\
Members with period& 33+(1) & 49+(0) & 54+(0) & 44+(0) & 24+(0) & 204+(1)\\
Members without period & 17+(1) & 11+(1) & 27+(0) & 29+(1) & 43+(1) & 127+(4)\\
Fraction members with period (c)& 0.55 & 0.69 & 0.57 & 0.52 & 0.30 & 0.50\\
Fraction with $N_R \leq 0.1$ & 0.55 & 0.86 & 1.00  & 1.00 & 1.00 & 0.89 \\\hline
\multicolumn{7}{l}{(a) Number of targets with  with SNR $\geq$5 where the number in brackets counts targets with unresolved membership.}\\
\multicolumn{6}{l}{(b) Percentage calculated neglecting targets with unresolved membership.}\\
\multicolumn{6}{l}{(c) Fraction after correction is made for the bias in the number of targets with resolved period in the spectroscopic sample.}\\
\end{tabular}
\end{table*}

Cluster members were iteratively defined as those stars with
RV less than $2\sigma_e$ from their mean RV, where $\sigma_e$ is
the effective velocity dispersion due to the combined effects of RV
uncertainties and the true velocity dispersion of the
cluster. The true velocity dispersion of the cluster was estimated
from the standard deviation of relative RV for targets with $v \sin i <20$\,km\,s$^{-1}$,
clipped at $\pm$5~km\,s$^{-1}$, giving a result of 0.66$\pm$0.17~km\,s$^{-1}$.
Hence, from equation~\ref{uncertainty_RV}

\begin{equation}
\sigma _e = \sqrt{(9.2+0.013(v \sin i)^2)^2/{\rm SNR}^2 + 0.72^2} \ \ {\rm km\,s}^{-1}.
\label{effective_uncertainty}
\end{equation}

Using equation~\ref{effective_uncertainty} yields 354 probable cluster
members, of which 210 have reported periods (see Fig.~2). A few fast
rotating stars with correspondingly large RV uncertainties were
included as members even though their RVs are some distance from the
mean in absolute terms. These are still very probable members since
fast rotators are comparatively rare amongst late-K to mid-M dwarf
field stars (Delfosse et al. 1998). To estimate the maximum number of
background stars falsely classified as members, the average number of
targets in $\pm 2\sigma_e$ bins centered at $\pm 10$~km\,s$^{-1}$ from the
cluster mean was counted. This indicates that of stars identified
as members, only 3/210 (1.5 per cent) with measured periods are
likely to be non members whereas about 12/144 (8 per cent)
without periods are possible non-members.

\subsection{Numbers of targets with and without measured periods}
In the following analysis the sample is restricted to 331 members with
$14.8 \leq I \leq 18.2$ (204 with periods, 127 without), where there is
sufficient target density (per unit magnitude) to make meaningful
comparisons of stars with and without periods. Table 5 shows the
numbers of stars with and without measured periods in five equally
spaced bins in $I$. Also shown in brackets are the numbers of stars in
each bin with SNR$\geq 5$~ but unresolved RVs. (see Table 2).

The fraction of stars with measured periods in our sample was biased by
our prior selection of a high proportion of targets with measured
rotation periods. In contrast, the fraction of photometric candidates
from IHA07 with measured periods is unbiased. The ratios of these
proportions are used to scale the numbers of non members in our sample
to obtain an unbiased estimate of the fraction of NGC~2516 members with
measured periods. Whilst the average is about 50 per cent, this falls
significantly for Bin 5 containing cooler stars. Finally, Table 5 shows
the fraction of targets in each bin with measured periods that also
have Rossby numbers $\leq 0.1$. These are expected to show saturated
chromospheric activity (see JJ10). A significant proportion (45 per
cent) of the stars in Bin 1 are unsaturated. These show an average
Ca\,{\sc ii}~IRT activity index of about half the saturated level. For
practical purposes all targets in the fainter (cooler) bins have
saturated chromospheric activity levels.

\section{The properties of low-mass stars with and without measured periods}
In this section we examine NGC~2516 members with
$6.7 \leq M_I \leq 10.1$ for evidence of differences in
the photometric, chromospheric or rotational properties of stars 
with and without measured periods.

\begin{figure*}
	\begin{minipage}[b]{0.85\textwidth}
	\centering
		\includegraphics[width=110mm]{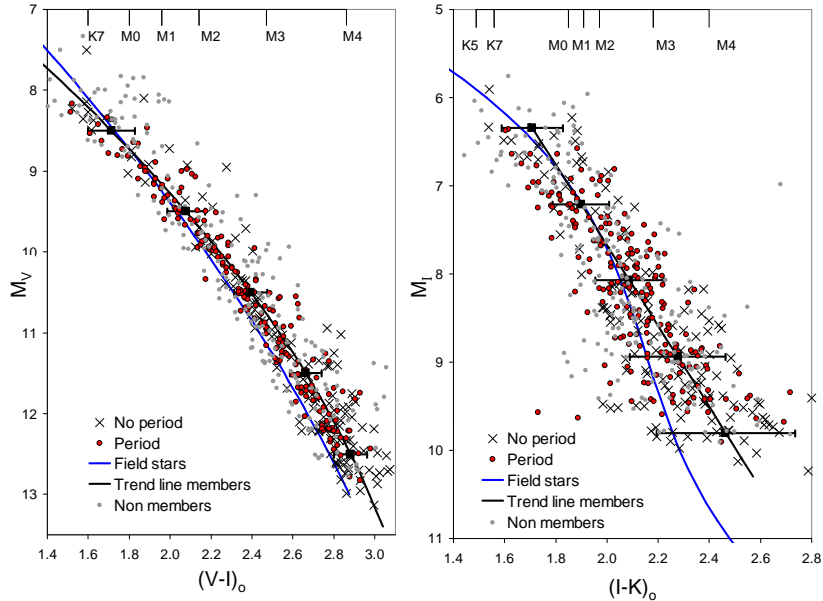}
\caption[$M_V$ versus $(V-I)_0$ and $M_I$ vs $(I-K)_0$ for members of
NGC~2516]{$M_V$ versus $(V-I)_0$ and $M_I$ vs $(I-K)_0$ colour
magnitude diagrams for targets in NGC~2516 with SNR$\geq5$~(using
$(M-m)_0=7.93$, $A_V=0.37$, $A_I=0.20$ and $A_K=0.04$). Circles and
crosses indicate targets confirmed as members (with and without
periods). Dots are
non-members (assuming the same distance modulus and
reddening). Also shown are: (a) a trend line 
to all cluster members together with 1$\sigma$ standard deviations;
(b) a polynomial representing the mean absolute
magnitude of nearby field stars as a function of colour (see section
4.1).}
	 \label{fig3}
	\end{minipage}
\end{figure*}

\subsection{Colour-magnitude diagrams}

Figure~5 shows $M_V$ versus $(V-I)_0$ and $M_I$ vs $(I-K)_0$ CMDs for
targets in NGC~2516 with measured RV and SNR $\geq$ 5.  Approximate
spectral types are indicated using colour to spectral type relations
from Kenyon and Hartmann (1995). Cluster members with and without
measured periods and RV non-members are indicated separately. Also
shown are trend lines representing a third order polynomial fit to all
cluster members together with 1$\sigma$ standard deviations and 
polynomials representing colour-magnitude relations for nearby field
stars. The latter are fits to field star data from the on-line database
of Reid (2002). Both plots show offsets between NGC~2516 members
and the trend line for field stars. This is not surprising since the
field stars are an older population with inhomogeneous metallicity. Of
more interest are the very small differences in the offsets for stars with
and without measured periods.

Stars with $6.7 < M_I < 10.1$ and measured periods have an offset of
$-0.012\pm 0.005$~mag in $(V-I)_0$ relative to the mean trend line,
whereas stars without measured periods have an offset of $0.009\pm
0.009$~mag. In the $M_I$ versus $(I-K )_0$ CMD stars with measured
period have an offset of $-0.006\pm 0. 008$~mag in $(I-K)_0$, whereas
stars without measured periods have an offset of $-0.007\pm 0.014$~mag.
Given that some small systematic differences between the two groups are
expected, for example in the expected level of field star contamination
in the two groups, these results do not indicate any difference in
colour between the two groups. By inference, there is little or no
difference in effective temperature as a function of luminosity between
stars with and without measured periods whereas one would be expected
if they had differing levels of starspot coverage (Jackson et
al. 2009).

\subsection{Chromospheric activity}
Figure 4 shows the Ca\,{\sc ii}~IRT activity indices for members of
NGC~2516 with and without measured periods as a function of absolute
$I$. Ca\,{\sc ii}~IRT activity was compared between the two data
sets by selecting cluster members with 6.7$<M_I <$10.1 and Ca\,{\sc
ii}~IRT activity index $>10^{-5}$ and measuring the offset of the
measured Ca\,{\sc ii}~IRT activity index from a linear regression line
as a function of $I$ for the whole dataset (see Fig.~4). The mean
offsets, -0.010$\pm$0.008~dex for targets with periods and
0.018$\pm$0.012~dex for members without periods, indicate no resolvable
difference in their mean levels of chromospheric
activity.

\subsection{Equatorial velocity}
The distribution of $v \sin i$ for M-dwarfs in NGC~2516 is similar to
that seen in other clusters of similar age (e.g. the Pleiades;
Queloz et al. 1998, Terndrup et al. 2000). There is a wide range of
rotation rates, with significant populations of more slowly rotating
stars ($v \sin i$ unresolved), a central band where the number density
falls with increasing $v \sin i$ with a scale length of about 30~km\,s$^{-1}$
and a tail of fast rotators with $50< v \sin i < 150$\,km\,s$^{-1}$.

\begin{figure}
\centering
\includegraphics [width = 80mm]{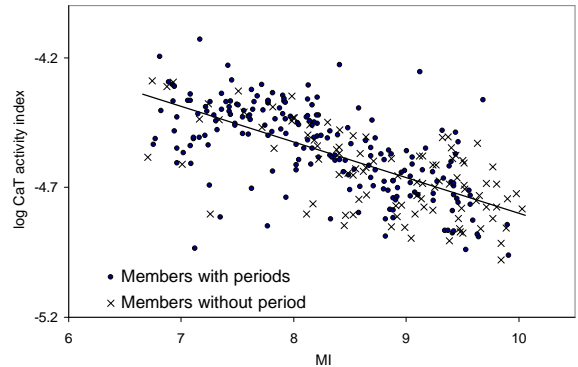}
\caption[Variation of chromospheric activity index with absolute I magnitude]
{Summed chromospheric activity indices for the first two lines
of the Ca\,{\sc ii}~IRT versus absolute $I$ for members of
NGC~2516 with spectral SNR $\geq$5 and a mean activity index
$>10^{-5}$. Also shown is a regression line for all members with
6.7$<M_I<$10.1.}
\label{fig4}
\end{figure}

The top left plot of Fig. 5 shows cumulative probability distributions
of $v\sin i$ for stars with $v\sin i \geq 8$\,km\,s$^{-1}$, with and
without measured periods. The distributions appear similar although
targets with measured periods have an excess tail of fast
rotators. A 2-sided Kolmogorov-Smirnov test yields a 75$\%$ probability
that the individual distributions are consistent with being drawn
from an identical distribution. The fractions of stars with $v\sin
i<8$\,km\,s$^{-1}$ (51/204 versus 26/127) are also indistinguishable.

Fig.~5 provides more
detailed comparisons of the $v\sin i$ distributions for members with
and without periods as a function of absolute $I$. 
The four bins containing higher mass stars
have between a 29 and $>$99 per cent probability that the
individual distributions are consistent with being drawn from the same
parent distribution. However, in Bin 5 the proportion of stars with
measured periods is reduced and the median $v\sin i$ is 7~km\,s$^{-1}$
higher for stars with measured periods than for those without
periods. This could signify fundamental differences in the velocity
distributions of cooler stars with and without the spotted
photospheres required to modulate light curves, but equally it could
reflect an anticipated measurement bias because it becomes increasingly
difficult to measure longer periods for fainter stars, due to
their larger photometric uncertainties (see Fig.~5 of IHA07).

\begin{figure*}
	\begin{minipage}[b]{0.85\textwidth}
  \centering
	\includegraphics[width = 110mm]{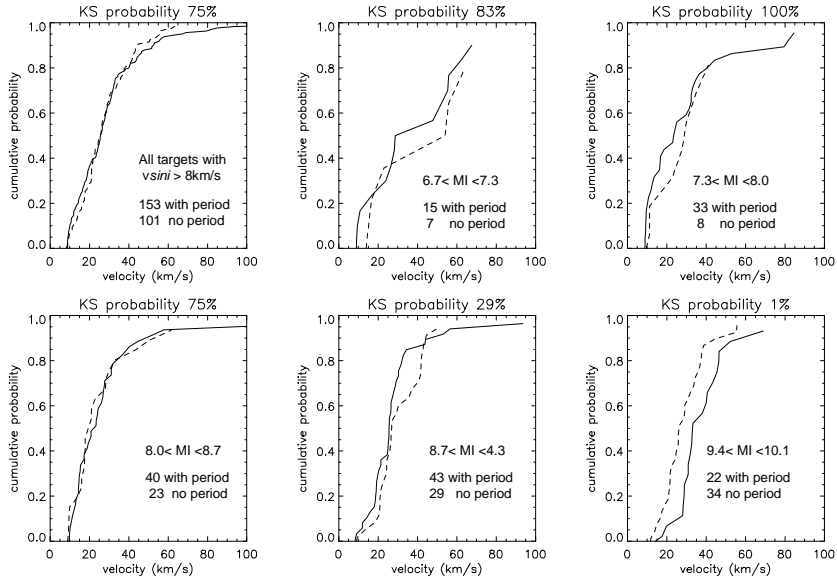}
	\caption[Comparison of the cumulative distribution functions of $v\sin i$]
	{A comparison of the cumulative distribution of $v\sin
	i$ for members of NGC~2516 with and without measured
	periods. The solid line shows stars with periods the dashed
	line shows those without. The top left plot compares all
	targets with $v\sin i\geq 8$~km\,s$^{-1}$. The remaining plots
	compare distributions in 5 equal bins of $I$ magnitude (see
	Table~5). Probabilities that data sets are consistent with
	being drawn from a common distribution are shown above each
	plot.}
	 \label{fig5}
	\end{minipage}
\end{figure*}

\subsection{A hypothesis based on the distribution of starspots}

If there are no significant differences in the colours, magnetic
activity and (for all but the faintest stars) no difference in rotation
speed for stars with and without measured periods, then what is causing
the apparent difference in their light curves?  An obvious possibility
is that there is some difference in observational 
sensitivity between the two groups.  Fig.~6 shows the expected light curve
amplitude uncertainty as a function of $I$. The majority of targets
were observed 270 to 320 times. Stars with and without measured periods
show similar levels of uncertainty. Thus accuracy of the observations
is not a discriminating factor between the two groups. Seven stars were
observed less than 270 times and show anomalously high uncertainties
(see Table 6). None have measured periods, but 6 of the 7 are bright
stars where only a small number of points were lost due to
saturation. For these stars, although the uncertainty is increased
above the norm, it could still be low enough to detect significant light
curve modulation.  One target (N2516-3-5-3994) was at
the extreme edge of a detector and was rarely sampled.  For simplicity
we decided to retain all these stars in our analysis.  As a check, the
comparisons of colour, chromospheric activity and the distribution of
$v\sin i$ for stars with and without measured periods were repeated
without these 7 stars.  The results were unchanged.

A second possible cause of observational bias is contamination of light
curves by adjacent stars. This was checked by comparing the frequency
with which a deblending algorithm was triggered during the photometric
analysis. The criterion for this was that isophotes of any adjacent
stars (averaged over 20 images) overlapped beyond a detection threshold
of 3 sigma above sky (Irwin 1985). The algorithm was triggered for 22
per cent of stars with measured periods compared to 28 per cent of
stars without periods. This difference is less than the Poisson
uncertainty for samples of 204 and 127 stars, so this appears not
to be a significant factor preventing detection of periods in one
group.

Having ruled out observational bias we can consider the physical
properties of stars in the two groups. Stars with measured periods
could show increased levels of spot coverage due to increased levels of
magnetic activity. The evidence is against this since the two groups
have indistinguishable levels of chromospheric (Ca\,{\sc ii}~IRT)
activity and are not displaced with respect to each other in CMDs. The
similar distributions of $v\sin i$ also suggest similar levels of
magnetic activity. It is possible that there is no relationship between
levels of chromospheric activity and spot coverage, in which case the
similarity of chromospheric activity levels has little significance,
but in our view this is unlikely. The flux of equipartion magnetic
field in M-dwarfs appears correlated with both rotation and magnetic
activity assessed with chromospheric and coronal indicators (Reiners,
Basri \& Browning 2009).

A second possibility is that stars without measured periods are slower
rotators and their rotational periods are too long to have been
measured by IHA07. This cannot be true since 80 per cent of members
without periods have $v\sin i \geq 8$~km\,s$^{-1}$ . There could be a
difference in the period distribution of the slower rotators (with
$v\sin i <8$~km\,s$^{-1}$).  However even this is not feasible since
the Monitor Survey of NGC~2516 claimed completeness $>75$ per cent for
periods $\leq 20$ days. Empirical activity-rotation relationships show
that saturated levels of magnetic activity in early- and mid-M dwarfs
occur for rotation periods shorter than 8--15 days and decrease rapidly
for longer periods (e.g. Kiraga \& Stepien 2007; Jeffries et
al. 2011). Hence, if there were a much higher proportion of long period
stars ($>20$ days) in the group without periods this would be indicated by
reduced chromospheric activity levels.  

A third possibility is
that stochastic variations associated with flaring activity might mask
starspot modulation. This seems unlikely because the rms variation in
the light curves of those stars without detected periods is
significantly lower on average than those in which modulation was
detected (see Table 7). That is, the light curves of stars without
detected periods are flatter, not noisier.



A final possibility is that the stars without measured periods are
predominantly viewed at low inclination angles, reducing their
light curve amplitudes. There are two arguments against this hypothesis.
\begin{enumerate}
	\item The inclination below which spot modulation becomes
	undetectable would have to be unreasonably high. In a simple
	model where half of stars show no modulation because of
	their inclination, the critical angle would be $i \simeq
	60^{\circ}$. Because the number density of objects with inclination
	$i$ varies as $\sin i$, most stars with undetected periods
	would have $40^{\circ} < i < 60^{\circ} $ but it seems unlikely this would
	be low enough to prevent spot modulation. There are many
	examples in the literature where spot modulation is seen in
	stars with $i <60^{\circ}$ (e.g LO Peg with $i=50^{\circ}$, Jeffries et
	al. 1994; HR 1099 with $i=38^{\circ}$, Lanza et al. 2006; PW And with
	$i=46^{\circ}$, Strassmeier and Rice 2006).

  \item A systematically lower inclination would lead to very different
  $v\sin i$ distributions in stars with and without periods. Making the
  reasonable assumption that rotational period and inclination are
  unrelated, then for a critical angle of $60^{\circ}$ the ratio of the
  mean value of $v\sin i$ for stars without periods to those with
  periods would be,

\begin{equation}
\int ^{\pi/3} _0 \sin^2  i\, di / \int ^{\pi/2} _{\pi/3} \sin^2 i \, di 	= 0.64
\end{equation}
\noindent{The similar $v\sin i$ distributions in Fig.~5 suggest this is
not the case. To test the significance of this consider
the $v\sin i$ distribution of the 204 stars with measured
periods, where 50 per cent have $v\sin i >$18.9\,km\,s$^{-1}$. If the
average $\sin i$ were reduced by a factor 0.64, then for the
same velocity distribution only 27 per cent of stars would have $v\sin
i$ above this value. Hence among the 127 stars without periods, we
would expect only 34 to have $v\sin i >$18.9\,km\,s$^{-1}$, compared
with the 77 observed. This discrepancy means that the possibility that
the difference between stars with and without measured periods is
determined solely by their inclination can be rejected with
$>99.9$~percent confidence according to a two-tailed chi-squared test.}

\end{enumerate}

\begin{figure}
	\centering
		\includegraphics[width = 84mm]{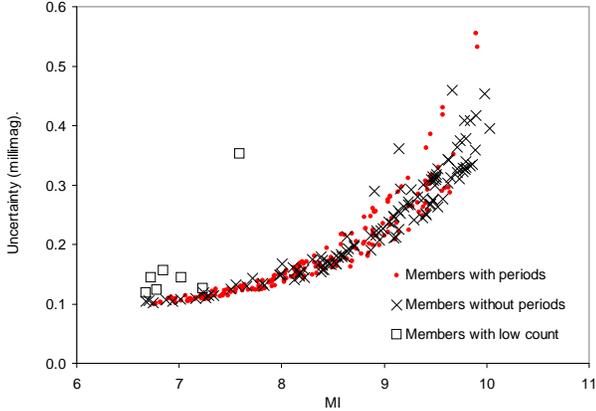}
	\caption{Predicted uncertainty in light curve amplitude as a
	function of $I$  ($6.7 < M_I < 10.1$) for members of
	NGC~2516 with and without periods. Circles and crosses show
	results for cluster members with $>270$ photometric 
        points defining their light
	curves. Squares show targets with lower numbers of points in
	their light curves.}
	\label{fig6}
\end{figure}

\begin{table}
	\caption{Targets with less likelihood of measuring a rotation period
	because there were fewer photometric
	points in their light curves.
	None of these targets have measured periods.}
\begin{center}
\begin{tabular}{lccc}	
\hline
Name &	M$_I$  &	v$\sin i$~(km\,s$^{-1}$) & Count\\
\hline
N2516-1-3-1740	& 6.68	 & -- & 234\\
N2516-1-8-3126	& 6.73	 & 13.1 & 155\\
N2516-3-5-3341	& 6.79	 & -- & 211\\
N2516-1-4-470	& 6.85	 &	-- & 141\\
N2516-1-4-401	& 7.03	&  23.0 & 181\\
N2516-1-4-673	& 7.24	 & -- & 244\\
N2516-3-5-3994	& 7.60	& 27.7 & 38\\\hline
\end{tabular}
\end{center}

\label{tab:Low counts}
\end{table}

Rejection of all these possibilities favours  hypotheses that do not 
require any fundamental difference between the two groups of stars.
If stars with and without measured periods have
similar distributions of rotation rate,
and similar levels magnetic activity and spot coverage then the
principle difference between stars with and without measured periods must
be a chance combination of:
\begin{enumerate}
	\item the distribution of spots on the stellar surface which
	affects the ``true'' light curve amplitude. This could be a
	random effect of starspot placement or a systematic effect where,
	for example, stars with measured periods have a higher proportion of
	equatorial spots.
	\item the effects of inclination and photometric uncertainties modulating the ``true'' light curve and producing the measured light curve from which a period is or is not measured.   
\end{enumerate}

\subsection{The distribution of light curve modulation amplitudes}

The method used by IHA07 to detect periodic modulation and determine
the periods and amplitudes was described in detail by Irwin et
al. (2006). Sine curves were fitted to data over a grid of
frequencies. Periodic light curves were selected on the basis of a
reduced $\chi^2$ test, followed by visual inspection of the folded
light curve. For stars with resolved periods IHA07 reported the first
harmonic amplitude of the best fitting sine wave. Results for our list
of target stars are reproduced in Table 4. Fig.~7 shows a histogram of
light curve amplitudes for the members of NGC~2516 with measured
periods. The most striking feature is how small the amplitudes are; the
mean is 0.016 mag. and all are below 0.05 mag.
\begin{figure}
\centering
	\includegraphics[width = 80mm]{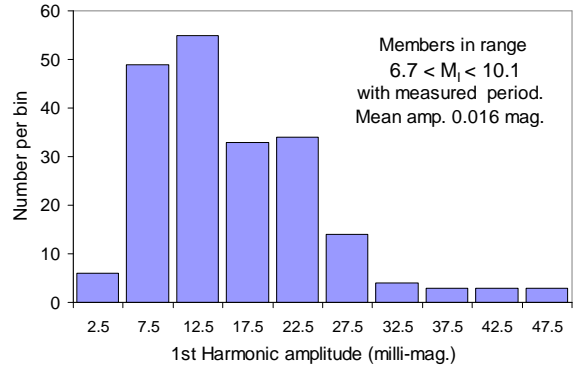}
	\caption{Histogram of light curve modulation amplitudes for targets with resolved periods and identified as members of NGC~2516.}
	\label{fig7}
\end{figure}

Using the fraction of stars
with measured periods from Table~5, the light curve amplitude data are
presented in Fig.~8 as probability distributions for the same five
$I$ magnitude bins considered in previous sections. The properties of
these distributions are summarised in Table~7.
Summing the probability densities for each dataset
gives the total probability of measuring a period in each bin.
The distribution functions for each bin are similar,
rising with harmonic amplitude before decaying to zero for
amplitudes $\geq0.05$~mag. The peak in probability density
shifts to higher amplitudes for the fainter stars of bins 4 and 5. This
could be a real effect or, more likely, reflects the increasing
difficulty in resolving periodicity for fainter stars, as
photometric uncertainties become comparable with light curve
amplitudes.
 
\begin{figure}
\centering
	\includegraphics[width = 80mm]{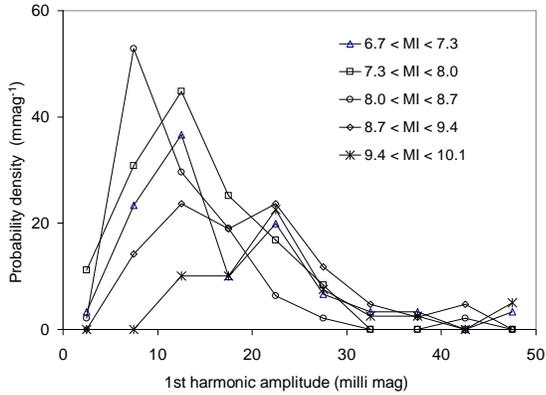}
	\caption{Probability density of light curve modulation
	amplitudes for members of NGC~2516 split according to their
	absolute $I$ magnitudes. The neasured frequencies are normalised
	according to the total number of targets in each set (see
	Table~5) such that the probability density summed over all
	amplitudes gives the fraction of targets in a set with measured
	period.}
	\label{fig8}
\end{figure}

\begin{table*}
\caption{Average characteristics of the $I$ band light curves of
  low-mass members of NGC~2516 as a function of absolute $I$ magnitude}
\begin{tabular}{lcccccc}
\\\hline
$M_I$ & 6.7 to 7.3 & 7.3 to 8.0 & 8.0 to 8.7 & 8.7 to 9.4
 & 9.38 to 10.06 & 6.66 to 10.06 \\\hline
Light curve rms for members with period (mag) & 0.010  & 0.008   & 0.008 & 0.012 & 0.015 & 0.010\\
Light curve rms for members without period (mag) & 0.007  & 0.003   & 0.005 & 0.008 & 0.006 & 0.008\\
Mean first harmonic of light curve (mag o-p) (a) &0.017 & 0.014 & 0.012 & 0.020  & 0.024 & 0.016 \\
Maximum light curve amplitude (mag o-p) (a) &0.049 & 0.030 & 0.041 & 0.045 & 0.049 & 0.049\\\hline
\multicolumn{7}{l}{(a) Stars with measured periods}\\

\end{tabular}
\end{table*}

\section{Discussion}
A previous analysis of a subset of these data used $v\sin i$ and
rotational period to estimate projected radii ($R\sin i$) for 210 fast
rotating members of NGC~2516 and determine mean stellar radii (Jackson
et al. 2009).  Stars with $M_I > 7.3$ have measured radii much higher
than expected from evolutionary models and much higher than inactive
field stars of similar luminosity. The larger radii and cooler
effective temperatures are consistent with up to $\sim$50~percent of
the stellar surface being covered by dark starspots (see Fig.~3
of Jackson et al. 2009). Despite this large starspot coverage these
stars show only low levels of variation in luminosity with rotation
(see Table 7 and Fig.~8), indicating that only a small fraction of the
darkened surface is modulating the light curve. In quantitative terms,
a single dark spot covering 1 per cent of the star's total surface is
sufficient to cause the typical light curve amplitudes of
0.02\,mag. This suggests as little as $\sim$ 2 per cent of the darkened
area is responsible for the light curve amplitude.

A possible explanation could be that spots are concentrated at
uneclipsed polar latitudes.  This effect would have to
be extreme; up to 98 per cent of the darkened surface would need to be
in axisymmetric spots or permanently visible (or invisible) 
at the poles. Young et
al. (1990) proposed high latitude activity to explain the light curves
of a nearby, rapidly rotating M0 dwarf, HK~Aqr.  However, more recent
Doppler imaging measurements of this and a second rapidly rotating M1.5
dwarf (EY Dra) by Barnes \& Collier Cameron (2001) showed no evidence
for polar spots.  The majority of spots appeared at low latitudes on HK
Aqr. EY Dra showed spots at all latitudes. Evidence for spot
concentration, polar or equatorial, remains sparse for late K and early
M-dwarfs with few measured in any detail (e.g. Barnes et
al. 2005). There is no evidence of polar spots extending down to
the latitudes of $30^{\circ}$ that would be required to cover 50 per cent of
the stellar surface.

An alternative explanation is that a large number of small starspots
are randomly distributed over the stellar surface. As the star rotates
the light curve would correspond to the average luminosity of many
small areas of varying luminosity. If these areas are sufficiently
small then the variation in the stellar luminosity would be a small
fraction of the change in luminosity from the immaculate surface to the
starspot.  For example, consider a star with 50 per cent surface
coverage by 5000 randomly placed dark starspots of order $2^{\circ}$ in
diameter. At any time 2500 spots would be visible to the observer and
the variation of luminosity with rotation would be of the order
$1/\sqrt{2500}$ -- about $\sim$0.02\,mag as found in our sample (see
Table 7).

The same hyothesis also explains why some stars have measured periods
and some do not. There is no difference between the two groups of stars
it is simply a matter of chance. In some cases the variations in
luminosity over a cycle contain a distinct first harmonic component
producing a measured period. In other cases the first harmonic
component is so low it cannot be resolved from the background of higher
order harmonics in the light curve.

The random spot distribution hypothesis need not be the sole
explanation for low light curve amplitudes. 
There could effectively be two starspot distributions, one axisymmetric/polar
and one randomly distributed over the stellar surface. Both
distributions would reduce $T_{\rm eff}$ and increase the radius of
active stars (at a given luminosity) but only the randomly distributed
spots could contribute to the measured light curve modulation. However
any axisymmetric/polar component would have to be very large to
significantly change the starspot size required to produce the measured
light curve amplitudes, since this only varies as the square root of filling
factor. If half the spot coverage were permanently axisymmetric, this
would only increase the estimated size for the randomly distributed
spots from $\sim 2^{\circ}$ to $\sim 3^{\circ}$. A further possibility is the
presence of bright plages contributing an additional source of
variation in surface luminosity. However, it is reasonable to suppose
that were plages present in significant numbers then their scale
length would also need to be comparable to that of the dark starspots,
otherwise they would increase the light curve amplitudes.

Our simplified calculation neglects the important effects of inclination,
limb darkening and how random intensity variations combine to produce a
measurable light curve amplitude. 
As such it probably overestimates the number of spots
and underestimates their size. Even so it serves to demonstrate that
the observed high spot filling factor (up to $\sim$50 per cent) and low levels
of light curve amplitudes ($\sim$ 0.02\,mag) of fast rotating M-dwarfs
in NGC~2516 are consistent with their surface being covered by a large
number of randomly distributed starspots of order $2^{\circ}$ in
diameter.

Small starspots can also explain why the total magnetic flux derived
from Stokes V measurements of active M-dwarfs is only 6--15 per cent of
the total magnetic flux implied by Stokes I measurements, indicating
that the magnetic flux on M-dwarfs is stored in small scale components
(Reiners \& Basri 2009).  The Stokes I component is found from Zeeman
broadening of magnetically sensitive lines produced by magnetic field
along the line of sight, irrespective of polarity (Johns-Krull and
Valentini 1996). As such, it is sensitive to the average magnetic field
and total filling factor of starspots, but gives no indication of their
scale length or distribution. Conversely, the circularly polarized
component of magnetic field, the Stokes V component, detects total
field along the line of sight taking account of its polarity. Periodic
modulation of the Zeeman signature due to variations in projected
velocity with time are used to determine the surface distribution of
magnetic field, or more precisely, to estimate the magnetic field
distribution in terms of the spherical harmonics of poloidal and
toroidal fields that best match the data (Donati et
al. 2006). The surface detail that can be resolved depends on the limit
of the spherical harmonic expansion.  The resolution increases with the
rotational speed of the star but is typically 8 terms for moderate
rotators, limiting resolution to magnetic features of size $>7^{\circ}$
($180/8\pi$).  Morin et al. (2008) suggest a higher resolution
is possible (70 elements around the equatorial ring) but used the same
limit of 8 spherical harmonics for their measurements of mid-M dwarfs.

If a significant proportion of the surface of active M-dwarfs is
covered by starspots of scale length $\sim 2^{\circ}$, then these will
contribute strongly to the total Stokes I signature but will not be
resolved by Zeeman Doppler imaging and will not contribute
significantly to the Stokes V component, although there would be some
small signature at high filling factors whenever random groupings of
small starspots form larger features on the stellar surface. This
explanation depends critically on the actual scale length of starspots
being significantly less than the current angular resolution of Zeeman
Doppler imaging.

\section*{Summary}

In this paper we have presented intermediate resolution spectroscopic
observations for two large samples of low-mass stars in the young
cluster NGC 2516. The two samples are distinguished by whether previous
photometric observations were able to discern significant rotational
modulation in their $I$-band light curves. Cluster membership has been
confirmed using a common radial velocity criterion and after correcting
for target selection biases, we find that approximately 50 per cent of
low-mass NGC~2516 members fall into each category.

The projected equatorial velocities, positions in colour-magnitude
diagrams and levels of chromospheric activity for the two samples have
been compared and show no significant differences. The vast majority of
our sample show very high levels of magnetic activity and fast rotation
rates, whether they have measured rotation periods or not. Neither are
there any differences in observation cadence or sensitivity that would
explain why a large fraction of stars do not show spot-modulated light
curves.  Explanations involving differences in magnetic activity,
rotation rate or spin-axis inclination are ruled out and we argue that
the overall spot coverage is likely to be similar in both sets of
stars. Instead we propose that the lack of spot modulation in many
stars and the low amplitude (0.01-0.02 mag) modulation seen even in those
stars which do have measured periods, can be explained if the spot
coverage consists of large numbers of randomly placed spots of
diameter $\sim 2^{\circ}$.  Such spots would be impossible to resolve
using current Doppler imaging and Zeeman Doppler imaging techniques and
may explain why only a small fraction of magnetic flux is detected in
the large scale structures probed by Stokes V measurements.

The next step in testing this hypothesis will be to develop a
quantitative model examining spot coverage as a function of mass or
spectral-type and attempt to simultaneously explain the detailed
distribution of light curve modulation amplitudes and the fraction of
stars with and without measured rotational modulation.

\section*{acknowledgements}
Based on observations collected at the European Southern
Observatory, Paranal, Chile through observing programs 380.D-0479 and
266.D-5655. RJJ acknowledges receipt of a Wingate scholarship. The
authors would like to thank Dr J Irwin and Dr S Aigrain for their help
with target selection for this project and for providing additional,
previously unpublished, data relating to their photometric light
curves.

\nocite{Augeros2011a}
\nocite{Aigrain2007a}
\nocite{Bagnulo2003a}
\nocite{Baraffe1998a}
\nocite{Barnes2001a}
\nocite{Barnes2005a}
\nocite{Barnes2011a}
\nocite{Carpenter2001a}
\nocite{Chabrier2007a}
\nocite{Collier1994a}
\nocite{Cutri2003a}
\nocite{Czesla2009a}
\nocite{Dachs1989a}
\nocite{Delfosse1998a}
\nocite{Delorme2011a}
\nocite{Donati1997a}
\nocite{Donati2006a}
\nocite{Eaton1979a}
\nocite{Gray1984a}
\nocite{Hall1972a}
\nocite{Hartman2010a}
\nocite{Hodgkin2006a}
\nocite{Irwin1985a}
\nocite{Irwin2006a}
\nocite{Irwin2007a}
\nocite{Irwin2008b}
\nocite{Irwin2009a}
\nocite{Irwin2009b}
\nocite{Jackson2009a}
\nocite{Jackson2010b}
\nocite{James2010a}
\nocite{Jeffries1994a}
\nocite{Jeffries1998a}
\nocite{Jeffries2001a}
\nocite{Jeffries2011a}
\nocite{Johnskrull1996a}
\nocite{Johnskrull2000a}
\nocite{Kiraga2007a}
\nocite{Kenyon1995a}
\nocite{Lanza2006a}
\nocite{Lyra2006a}
\nocite{Makarov2009a}
\nocite{Mallik1994a}
\nocite{Mallik1997a}
\nocite{Marcy1982a}
\nocite{Marsden2009a}
\nocite{Meibom2011a}
\nocite{Messina2001a}
\nocite{Morales2010a}
\nocite{Moraux2005a}
\nocite{Morin2008a}
\nocite{Neff1995a}
\nocite{Noyes1984a}
\nocite{Odell1995a}
\nocite{Oneal2004a}
\nocite{Pickles1998a}
\nocite{Pizzolato2003a}
\nocite{Prosser1993a}
\nocite{Queloz1998a}
\nocite{Reid2002a}
\nocite{Reiners2009a}
\nocite{Reiners2009b}
\nocite{Ribas2008a}
\nocite{Rice1996a}
\nocite{Rieke1985a}
\nocite{Semel1989a}
\nocite{Strassmeier2002a}
\nocite{Strassmeier2006a}
\nocite{Strassmeier2009a}
\nocite{Sung2002a}
\nocite{Terndrup2000a}
\nocite{Terndrup2002a}
\nocite{Thomas2008a}
\nocite{vanLeeuwen1987a}
\nocite{Watson2004a}
\nocite{Vogt1981a}
\nocite{Young1990a}

\bibliographystyle{mn2e} 
\bibliography{References}


\bsp 

\label{lastpage}

\end{document}